\documentclass{aa}  

\usepackage{natbib}
\usepackage{graphicx}
\usepackage{txfonts}
\usepackage{hyperref}

\usepackage{bm}

\newcommand{\wmap}{\textsl{WMAP}}
\newcommand{\planck}{\textsl{Planck}}

\newcommand{\chandra}{\textsl{Chandra}}
\newcommand{\xmm}{\textsl{XMM-Newton}}

\newcommand{\beq}{\begin{equation}}
\newcommand{\eeq}{\end{equation}}
\newcommand{\beqa}{\begin{eqnarray}}
\newcommand{\eeqa}{\end{eqnarray}}
\newcommand{\mpc}{$h^{-1} \mathrm{Mpc}$}
\newcommand{\msun}{$h^{-1} {\rm M}_{\odot}$}
\newcommand{\ms}{${\rm M}_{\odot}$}

\newcommand{\eg}{e.g.,\xspace}

\def\der{{\rm d}}

\begin{document} 

  \title{Direct detection of the kinetic Sunyaev-Zel'dovich effect \\ in galaxy clusters}
  \author{Hideki Tanimura\inst{1} \and Saleem Zaroubi\inst{2,3,4} \and Nabila Aghanim\inst{1}}

  \institute{
    Universit\'{e} Paris-Saclay, CNRS, Institut d'Astrophysique Spatiale, B\^atiment 121, 91405 Orsay, France \and 
    Department of Natural Sciences, Open University of Israel, 1 University Road, PO Box 808, Ra'anana 4353701, Israel \and
    Kapteyn Astronomical Institute, University of Groningen, PO Box 800, NL-9700 AV Groningen, the Netherlands \and
    Department of Physics, The Technion, Haifa 32000, Israel \\
  \email{hideki.tanimura@ias.u-psud.fr}
             }
             
  \date{}

  \abstract 
    {We report the direct detection of the kinetic Sunyaev-Zel'dovich (kSZ) effect in galaxy clusters with a 3.5$\sigma$ significance level. The measurement was performed by stacking the \planck\ map at 217 GHz at the positions of galaxy clusters from the Wen-Han-Liu (WHL) catalog. To avoid the cancelation of positive and negative kSZ signals, we used the large-scale distribution of the Sloan Digital Sky Survey (SDSS) galaxies to estimate the peculiar velocities of the galaxy clusters along the line of sight and incorporated the sign in the velocity-weighted stacking of the kSZ signals.
    Using this technique, we were able to measure the kSZ signal around galaxy clusters beyond $3 \times R_{500}$. Assuming a standard $\beta$ -model, we also found that the gas fraction within $R_{500}$ is $f_{gas,500} = 0.12 \pm 0.04$  for the clusters with the mass of $M_{500} \sim 1.0 \times 10^{14}$ \msun.  We compared this result to predictions from the Magneticum cosmological hydrodynamic simulations as well as other kSZ and X-ray measurements, most of which show a lower gas fraction than the universal baryon fraction for the same mass of clusters. Our value is statistically consistent with results from the measurements and simulations and also with the universal value within our measurement uncertainty. 
    }
    \keywords{galaxies: clusters: general - intracluster medium - Cosmology:large-scale structure of Universe - cosmic background radiation}

\maketitle


\section{Introduction}
\label{sec:intro}

Galaxy clusters are the largest gravitationally bound structures in the Universe and have been used as probes of cosmology and astrophysics. These massive objects imprint their signature on the cosmic microwave background (CMB) through the Sunyaev-Zel'dovich (SZ) effect \citep{Sunyaev1970, Sunyaev1972, Sunyaev1980}. The SZ effect is caused by the scattering of CMB photons by hot, ionized plasma in the intracluster medium (ICM) giving rise to a change in the CMB temperature. The SZ effect can be classified into two contributions: the thermal Sunyaev-Zel'dovich (tSZ) effect and the kinetic Sunyaev-Zel'dovich (kSZ) effect (e.g \citealt{Diaferio2000}). 

The tSZ effect is caused by the scattering of CMB photons by free electrons with thermal motions in galaxy clusters, resulting in a characteristic spectral distortion of the CMB blackbody spectrum. This effect allows us to trace gas pressure in these objects, and has been well characterized through measurements of individual massive clusters (\eg \citealt{Plagge2010, Bonamente2012, Sayers2013a, Planck2013IRV, Romero2015}) and statistical measurements of massive to low-mass systems (e.g. \citealt{Planck2013IRXI, Greco2015, Vikram2017, Hill2018, Lim2018, Tanimura2019, Tanimura2020}). 

The kSZ effect, in turn, is caused by the scattering of CMB photons off the electrons due to the peculiar motion of a galaxy cluster, leading to a Doppler shift of the CMB blackbody spectrum. While kSZ effect is elusive due to its small amplitude and identical spectral shape to the CMB spectrum,  kSZ has great potential in constraining both cosmological and astrophysical models. 
From a cosmological point of view, peculiar velocities of galaxy clusters are found through measurements of the kSZ effect and they allow the amplitude of the growth rate of density fluctuations to be estimated. In coming years, it is expected that kSZ measurements by the advanced Atacama Cosmology Telescope (AdvACT) \citep{Henderson2016} and the third-generation South Pole Telescope  (SPT-3G) \citep{Benson2014} will constrain the growth rate of structures to sub-percentage precision at $z<1$ \citep{Alonso2016}, which then help to constrain models of dark energy \citep{Bhattachary2008,Ma2014}, modified gravity \citep{Mueller2015b, Bianchin2016}, and massive neutrinos \citep{Mueller2015a}. 
From an astrophysical point of view, there is debate over whether a significant fraction of diffuse gas is present around halos as a circumgalactic medium, or whether the gas, once expelled because of feedback processes such as star formation, supernovae, and active galactic nuclei (AGNs), is never accreted onto the halos (\eg \citealt{Planck2013IRXI, Anderson2015, Brun2015}). As the kSZ effect is sensitive to the virialized gas and also to the gas surrounding halos, independent of the gas temperature (unlike the tSZ effect), it is well suited to studying these baryons through the distribution of gas around galaxy clusters. 

The kSZ signal has so far been detected for a few individual systems (\eg \citealt{Sayers2013b, Adam2017}) or by statistical measurements based on the pairwise method (\eg \citealt{Hand2012, Hernandez2015, Planck2016IRXXXVII, Soergel2016, Bernardis2017}) or cross-correlation method \citep{Hill2016}. However, the reported significance of these detections is limited to $\sim$2--4$\sigma$. 
To extract the kSZ signal, most of these statistical measurements rely on a matched filter (\eg \citealt{Soergel2016, Lim2020}) or aperture photometry (\eg \citealt{Hernandez2015, Planck2016IRXXXVII, Schaan2016, Bernardis2017, Sugiyama2018}). However, due to the large uncertainty in the gas density profile in galaxy clusters, the kSZ signal extracted by a matched filter can be biased if the assumed profile is incorrect \citep{Ferraro2015}. Aperture photometry has less dependence on the density profile, however the extracted signal level changes depending on the aperture size because of the varying signal-to-noise ratio (S/N).  

In the present analysis, to extract the kSZ signal, we use a stacking method without making any assumptions regarding the spatial distribution of gas around galaxy clusters or setting any aperture. To avoid cancelation of the kSZ signals when all the clusters are simply stacked, we estimate the peculiar velocities of the galaxy clusters along the line of sight (LOS) with which the sign of the kSZ signals is attributed during the stacking. The paper is organized as follows. Section \ref{sec:data} summarizes data sets used in our analyses. Section \ref{sec:vel} explains the velocity reconstruction of galaxy clusters.  Section \ref{sec:ana} explains the stacking method to extract the kSZ signals. Possible systematic errors in our measurements are discussed in Section \ref{sec:systematics}. The interpretation of the measurements is presented in Section \ref{sec:interp}. We end this paper with discussions and conclusions in Section \ref{sec:discussion} and Section \ref{sec:conclusion}. 

Throughout this work, we adopt the $\Lambda$CDM cosmology in \cite{Komatsu2011} with $\Omega_{\rm m} = 0.272$, $\Omega_{\rm b} = 0.046$, and $H_0 = 70.4$ km s$^{-1}$ Mpc$^{-1}$. The cosmological parameters are used to estimate peculiar velocities of galaxy clusters and distances to them, and our result may depend on the assumed cosmological parameters. However, we perform our data analysis with \planck\ cosmology in \cite{Planck2016XIII} and obtain consistent results. All masses are quoted in units of solar mass, and $M_{\Delta}$ is mass enclosed within a sphere of radius $R_{\Delta}$ such that the enclosed density is $\Delta$ times the {\it critical} density at redshift $z$. Uncertainties are given at the 1$\sigma$ confidence level.


\section{Data}
\label{sec:data}

\subsection{Galaxy cluster catalog}
\label{subsec:whl}
\cite{Wen2012} and \cite{Wen2015} identified a total of 158,103 Wen-Han-Liu galaxy groups and clusters from the Sloan Digital Sky Survey (SDSS) galaxies in the redshift range between 0.05 and 0.8, of which 89\% have spectroscopic redshifts (hereafter WHL galaxy clusters). 
The masses of the WHL galaxy clusters were estimated from their total luminosity, and these were calibrated by the masses of 1191 clusters using X-ray or tSZ measurements. In our study, we use WHL galaxy clusters that have spectroscopic redshifts at 0.25<$z$<0.55 and masses of $M_{500} > 10^{13.5}$\msun. In addition, after selections in Sect. \ref{sec:vel} and Sect. \ref{subsec:stacking}, the total number of clusters used in our analysis is 30,431. 

\subsection{Galaxy catalog}
\label{subsec:sdss}
The SDSS galaxies in \cite{Reid2016} are composed of 953,193 galaxies in the northern Galactic hemisphere and of 372,542 in the southern Galactic hemisphere, combining the the Baryon Oscillation Spectroscopic Survey (BOSS) LOWZ galaxies and constant-mass (CMASS) galaxies galaxies. The completeness of the galaxies is stated to be 99\% for CMASS and 97\% for LOWZ. Spectroscopic information is available for all the galaxies and their redshifts extend up to z $\sim$ 0.8. 
We used the galaxies to compute the galaxy density field, which was then used to estimate peculiar velocities of the WHL clusters. To avoid a bias in the velocity calculation due to the magnitude limit at different redshifts, we limit our analysis to the range of 0.25<$z$<0.55, in which the number density of the galaxies in the survey volume is fairly flat (see Fig.~11 in \citealt{Reid2016}).

\subsection{Planck maps}
\label{subsec:planck}
\planck\ produced all-sky maps in nine frequency bands from 30 to 857 GHz with the angular resolutions ranging from 31' to 5'. In our study, for the detection of the kSZ signal, we mainly used the 217 GHz frequency map from the Planck 2018 data release \citep{Planck2018III}, because it corresponds to the null frequency of the tSZ effect. We also used, for comparison, the \planck\ CMB maps produced using four different component-separation methods \citep{Planck2018IV}: COMMANDER (Optimal Monte-carlo Markov chAiN Driven EstimatoR) \citep{Eriksen2004, Eriksen2008}, NILC (Needlet Internal Linear Combination) \citep{Delabrouille2009}, SEVEM (Spectral estimation via expectation maximisation) \citep{Martinez2003, Leach2008,  Fernandez2012}, and SMICA (Spectral Matching Independent Component Analysis) \citep{Delabrouille2003, Cardoso2008}. These maps were provided in HEALpix\footnote{http://healpix.sourceforge.net/} format \citep{gorski2005} with a pixel resolution of $N_{\rm side}$ = 2048 ($\sim 1.7$ arcmin). To minimize the Galactic and extragalactic contamination, we applied the mask produced by the \planck\ team for the analysis of the CMB temperature maps, which masks the region around the Galactic plane and the point sources detected at all the \planck\ frequencies (see Table C.1 in \citealt{Planck2018IV}). In addition, we used a more robust point-source mask, masking radio and infrared sources, which is used for the analysis of the Compton $y$ maps \citep{Planck2016XXII}. Combining these two masks excludes $\sim$50\% of the sky. 

\subsection{Magneticum simulation}
\label{subsec:magneticum}
The Magneticum simulations are one of the largest cosmological hydrodynamical simulations \citep{Hirschmann2014, Dolag2015}, and are based on the standard $\Lambda$CDM cosmology from \cite{Komatsu2011} with $\Omega_{\rm m} = 0.272$, $\Omega_{\rm b} = 0.046$, and $H_0 = 70.4$ km s$^{-1}$ Mpc$^{-1}$. They provide several public datasets\footnote{http://www.magneticum.org/data.html\#} \citep{Ragagnin2017}, of which we use the simulated kSZ light-cone map with an area of $\sim$1600 deg$^2$ (hereafter Magneticum light cone). The corresponding cluster catalog with $M_{500} > 3 \times 10^{13}$ \msun\ at $z \lesssim 2$ is also provided \citep{Dolag2016,Soergel2018}. In addition, they also provide the post-processed data of the galaxy catalog and the cluster catalog from the full simulations with a cube box of 640$^3$ \mpc. We use the post-processed data of ``Box2b\_hr'' at $z\sim$ 0.42 (hereafter Magneticum snapshot), corresponding to the median redshift in our analysis range.


\section{Velocity reconstruction}
\label{sec:vel}

The peculiar velocity of a galaxy cluster at the position, $\bf{x}$, can be derived in the linear regime by
\beq
    \bm{\varv}({\bf{x}}) = \frac{f(\Omega) a H(a)}{4 \pi} \int \der^3 \bf{y} \, \delta(\bf{y}) \, \frac{\bf{y} - \bf{x}}{|\bf{y} - \bf{x}|^3}, 
    \label{eq:vel}
\eeq
where $a$ is the scale factor, $H(a)$ is the Hubble parameter, $f(\Omega)$ is the linear velocity growth rate given by $f(\Omega) \simeq \Omega_{\rm m}^{0.545}$ \citep{Lahav1991, Wang1998, Linder2005, Huterer2007, Ferreira2010}, and $\delta(\bf{y})$ is the overdensity of matter at the position $\bm{y}$. 

To compute the matter density field, $\delta(\bf{y})$, we used the galaxies in Sect. \ref{subsec:sdss}. Their redshift distances and J2000.0 coordinates were transformed into the comoving Cartesian coordinates: 
\beqa
    X &=& r(z) \cos \alpha \cos \delta, \nonumber \\
    Y &=& r(z) \sin \alpha \cos \delta, \nonumber \\
    Z &=& r(z) \sin \delta,  
    \label{eq:cartesian}
\eeqa
where $\alpha$ and $\delta$ refer to the J2000,0 right ascension and declination, respectively, and $r(z)$ is the comoving distance at redshift $z$, with which a number density of galaxies was computed in the SDSS survey field. The galaxy density field can be connected to the matter density field by $\delta_{g} = b \, \delta$ through the galaxy bias, $b$. The galaxy bias was studied for the LOWZ galaxies \citep{Parejko2013} and CMASS galaxies \citep{White2011, Nuza2013, Rodriguez2016}, showing $b \sim 2$ at scales larger than 10 \mpc. Since a much larger scale of $\sim$240 \mpc, as later described, was considered in our study to calculate the peculiar velocity, we used $b = 2$. 
The estimated galaxy bias has a relatively large uncertainty at large scales, which is not included in our analysis. 
While we did not find a significant bias in the reconstructed peculiar velocities using $b = 2$ as shown in the following Fig.~\ref{fig:vel-comp},  a more precise estimate of galaxy bias would be needed for a more precise measurement of peculiar velocity. 

In practice, we placed a galaxy cluster at the center in a cubic box of 240$^3$ \mpc, in which the box was divided by grid cells of $5^{3}$ \mpc. 
If the cubic box for a galaxy cluster reaches the edge of the SDSS survey, we removed the galaxy cluster from our catalog because it may bias the velocity estimate. We then placed galaxies around the galaxy cluster in the box cells and calculated the overdensities of the galaxies. 
While finer grids may be better for the velocity estimate, we determined the size so that the grid size is large enough compared to the length expected from redshift-space distortion (RSD): The RSD distortion for a typical velocity of 300 km/s is $\sim$3 \mpc\ at the redshift of interest in our analysis.
Box cells were then smoothed by a Gaussian kernel of 2 \mpc\ to remove sharp grid edges. By applying Eq.\,\ref{eq:vel} to this cubic box, the peculiar velocity of a galaxy cluster was obtained. 

We checked our velocity estimates of galaxy clusters using the Magneticum snapshot simulation by comparing the velocities of the simulated galaxy clusters computed using Eq.\,\ref{eq:vel} with the true velocities. To reproduce ``real'' data in the simulation, we removed the simulated galaxy clusters with $M_{500} < 10^{13.5}$\msun\ (minimum mass of the WHL clusters we use), and also removed the simulated galaxies with $M_{*} < 1.8 \times 10^{11}$ \msun, so that the number density of galaxies is the same as the real data. In this simulation box, we calculated the 3D velocities of the simulated galaxy clusters and compared the velocities in the ``LOS'' direction, which was defined by placing an observer at the corner of the simulation box. From the comparison, we find that our velocity estimates are well correlated with the true ones as shown in the left panel in Fig.~\ref{fig:vel-comp} with an uncertainty of $\sim$180 km/s following a Gaussian distribution.

    \begin{figure}
    \centering
    \includegraphics[width=0.49\linewidth]{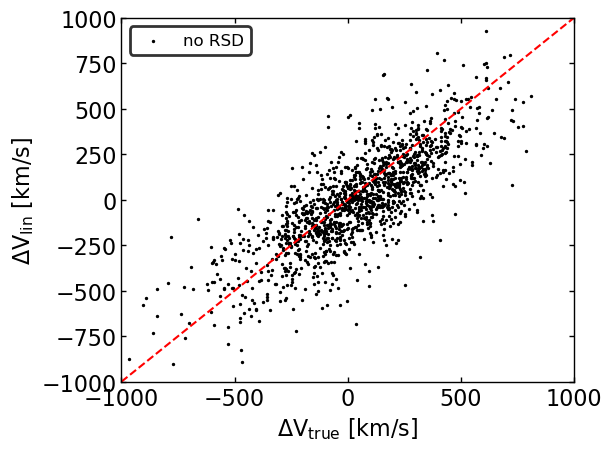}
    \includegraphics[width=0.49\linewidth]{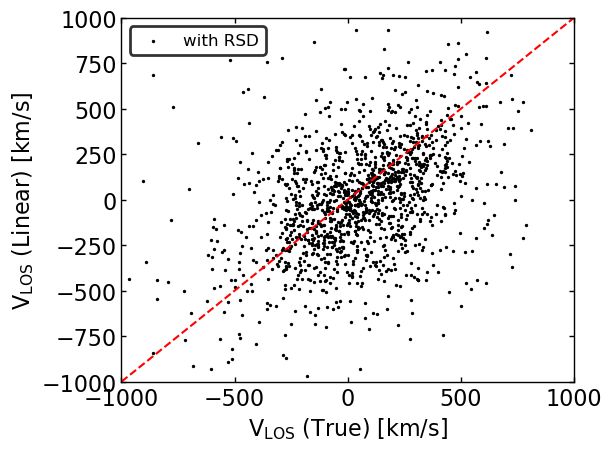}
    \caption{X-axis: True LOS velocities of simulated galaxy clusters in the Magneticum simulations. Y-axis: LOS velocities of the same galaxy clusters computed using Eq.\,\ref{eq:vel} with the galaxy bias of $b=2$ when the RSD effect is not included  {\it (left)} and included  {\it (right)}.}
    \label{fig:vel-comp}
    \end{figure}

In addition, we checked the effect of RSD on the velocity estimates using the Magneticum snapshot simulation by including the RSD in the simulated galaxies. For simplicity, we did not re-identify galaxy clusters in the redshift space, but used the same galaxy clusters with their positions redshifted. By again comparing the computed velocities of the simulated galaxy clusters with the true velocities, we find that our velocity estimates are well correlated with the true ones as shown in the right panel of Fig.~\ref{fig:vel-comp}, with a larger uncertainty of $\sim$260 km/s following a Gaussian distribution. We also find a negative bias of $\sim$40 km/s relative to the true values, but this bias is not significant given the current uncertainty.


\section{Analysis}
\label{sec:ana}

To detect the kSZ signal, we applied a stacking method for the \planck\ HFI 217 GHz map or \planck\ CMB maps. However, because there is equal probability that a cluster will have positive or negative LOS velocity, the associated kSZ signal from clusters cancels out by a simple stacking. Therefore, to avoid the cancelation of equally likely positive and negative kSZ signals, we first performed the stacking after separating galaxy clusters depending on the directions of the LOS velocities estimated in Sect. \ref{sec:vel}. 

In this paper, we define ``positive'' LOS direction as a radial direction from us: a positive motion is the motion moving away from us and a negative motion is the motion approaching us. It follows that when a galaxy cluster has a positive motion, the CMB is redshifted, resulting in a negative kSZ signal. On the other hand, when a galaxy cluster has a negative motion, the CMB is blueshifted, resulting in a positive kSZ signal.

The stacking was then also performed for each cluster weighted by the LOS velocity. A positive kSZ signal weighted by a negative LOS velocity then has a negative signal and a negative kSZ signal weighted by a positive LOS velocity also has a negative signal. This allows us to set the kSZ signal to a negative value, while other components are canceled out by the positive or negative LOS velocity with equal probability.

\subsection{Filtering the cosmic microwave background}
\begin{figure}
    \centering
    \includegraphics[width=\linewidth]{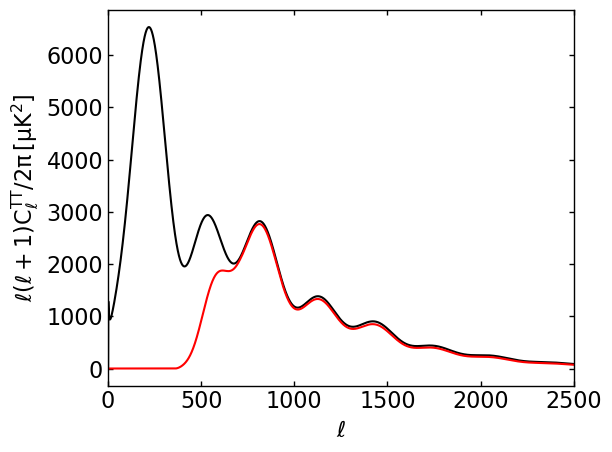}
    \caption{Filter applied to the \planck\ maps. The filter is a smooth function with its response of one below 15 arcmin ($\ell\sim$720) and zero above 30 arcmin ($\ell\sim$360). With this filter, the primordial CMB fluctuations in {\it black} are suppressed to the ones in {\it red}. }
    \label{fig:cmb-filter}
    \end{figure}

The amplitude of the kSZ signal around galaxy clusters is of the order of $\sim$1 uK, and is dominated by the primordial CMB fluctuations of the order of $\sim50$ uK. To extract the kSZ signal present at the cluster scales, that is, small angular scales, we applied a spatial filter to the \planck\ maps. The angular size of the virial radius of the WHL galaxy clusters is 2.2 -- 10.5 arcmin. We therefore filtered out angular scales above 15 arcmin ($\ell\sim 720$) in Fourier space using a smooth function with its response of one below 15 arcmin ($\ell\sim$720) and zero above 30 arcmin ($\ell\sim$360). The effect of the filter is shown in Fig.~\ref{fig:cmb-filter} for a simulated power spectrum of the primordial CMB fluctuations. With this filter, the standard deviation of the primordial CMB fluctuations is reduced to $\sim40$ uK.

\subsection{Stacking}
\label{subsec:stacking}
\begin{figure*}
    \centering
    \includegraphics[width=0.33\linewidth]{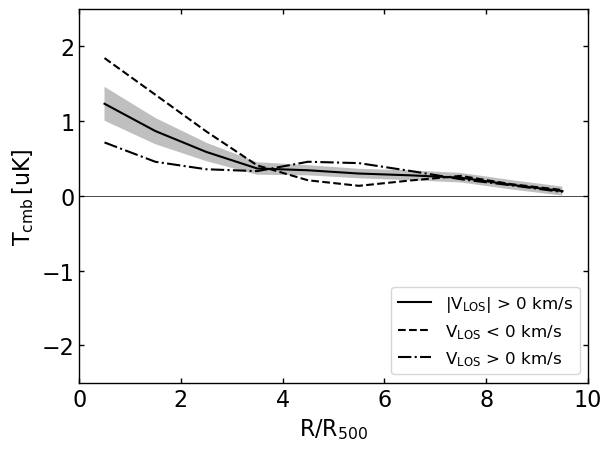}
    \includegraphics[width=0.33\linewidth]{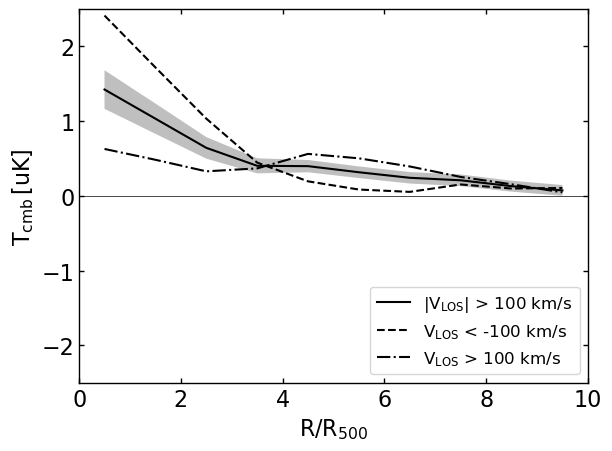}
    \includegraphics[width=0.33\linewidth]{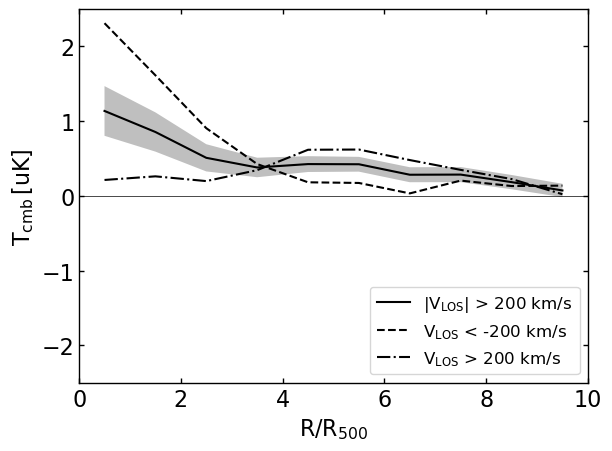}
    \includegraphics[width=0.33\linewidth]{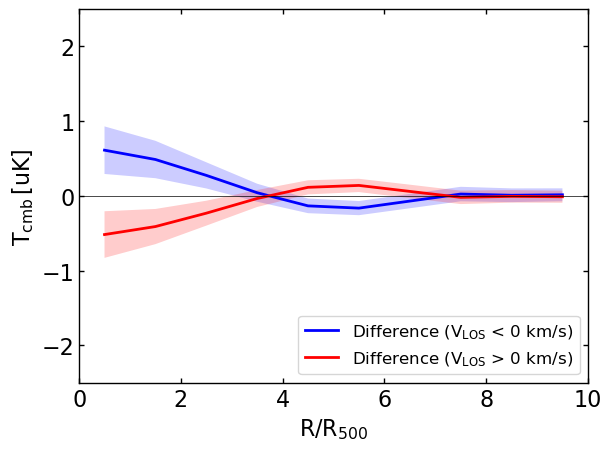}
    \includegraphics[width=0.33\linewidth]{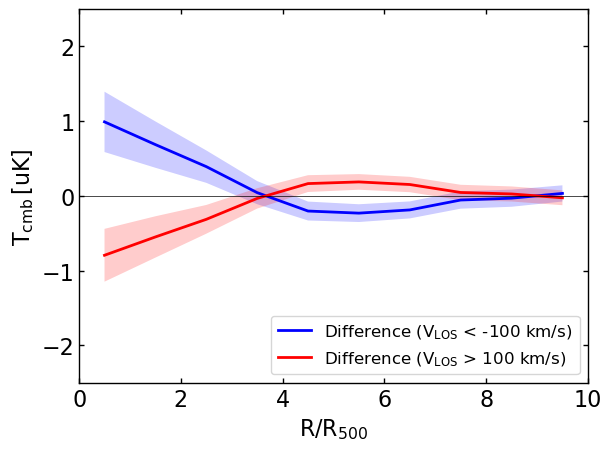}
    \includegraphics[width=0.33\linewidth]{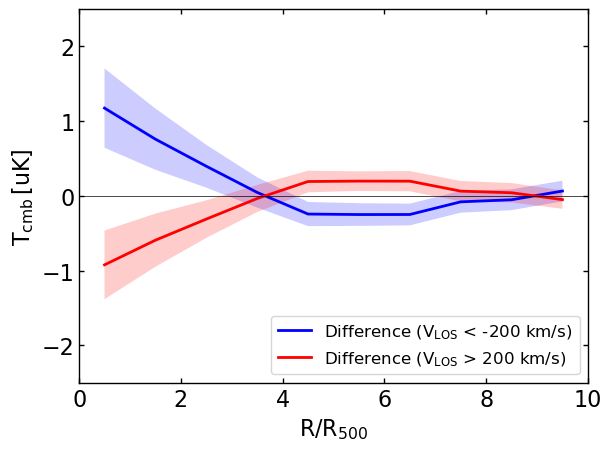}
    \caption{({\it Upper panels}) {\it Left}: Stacked radial profile around the 30,431 WHL clusters with the \planck\ temperature map at 217 GHz ({\it black}) with the 1$\sigma$ uncertainty estimated by a bootstrap resampling ({\it gray}). The stacking is performed separately for the 14,093 WHL clusters with negative LOS motions ({\it black dashed line}) and for the 16,338 WHL clusters with positive motions ({\it black dash-dotted line}). {\it Middle}: Same as {\it left} panel but with an additional LOS velocity cut of 100 km/s. {\it Right}: Same as {\it left} panel but with an additional LOS velocity cut of 200 km/s. ({\it Lower panels}) Positive kSZ radial profile ({\it blue}) extracted by taking the difference between the {\it black dashed} and {\it black} lines, and negative kSZ radial profile ({\it red}) extracted by taking the difference between the {\it black} and {\it black dash-dotted} lines.   }
    \label{fig:1stack-ksz}
    \end{figure*}

We stacked the {\it filtered} \planck\ maps at the positions of the WHL galaxy clusters and constructed the stacked radial profile. In practice, we placed each galaxy cluster at the center of a two-dimensional grid in ``scaled'' angular distance in the range of  $-10 < \theta/\theta_{500} < 10$, divided into 10 $\times$ 10 bins, where $\theta_{500}$ is the angular radius of a galaxy cluster calculated with $R_{500}$ provided in the catalog. 
The \planck\ maps were scaled accordingly and data were placed on the two-dimensional grids, while data in the masked region were not used. In this process, if more than 20\% of the region within $10 \times \theta_{500}$ around a galaxy cluster was masked, the galaxy cluster was removed from our catalog, because a large mask may bias our measured profile.  We repeated this for the selected 30,431 galaxy clusters and computed the stacked radial profile. 
In the upper left panel of Fig.~\ref{fig:1stack-ksz}, the stacked radial profile of the galaxy clusters using the \planck\ HFI 217 GHz map is shown in {\it black}.
The stacked profile includes all the Galactic and extragalactic emissions in the frequency in which the cosmic infrared background (CIB) is dominant at small scales as shown in Appendix \ref{sec:comp}. In this step, no velocity weighting is applied and hence positive and negative kSZ signals are canceled out.

To extract the kSZ signals of the galaxy clusters, we stacked the galaxy clusters with positive (16,338 clusters) and negative (14,093 clusters) LOS motions separately. 
As expected, the original stacked profile with all the galaxy clusters (30,431 clusters) was separated into two: one with stacked positive kSZ signal and the other with stacked negative kSZ signal as shown in Fig.~\ref{fig:1stack-ksz}. As, in these profiles, the same amount of Galactic and extragalactic components are included on average, we can extract the kSZ signals by taking the differences of the two separated profiles relative to the original one, shown in Fig.~\ref{fig:1stack-ksz}.

To confirm whether or not the separated signals originate from the kSZ, 
we checked the correlation between the amplitude of the separated signals and the LOS velocity, which is only expected for the kSZ but not for other components such as CMB, CIB, and tSZ. In the upper middle and right panel of Fig.~\ref{fig:1stack-ksz}, the stacking was performed for the galaxy clusters with the absolute LOS velocity larger than 100 km/s or 200 km/s, and the separated signals are shown in {\it blue and red} in the lower panels. The results show that the amplitude of the separated signals increases along with the amplitude of the velocity cut. This trend clearly supports that the separated signals originate from the kSZ. 

We assessed the uncertainties of the stacked profile through bootstrap resampling.  We drew a random sample of the galaxy clusters with replacement and re-calculated one stacked profile for the new set of galaxy clusters. We repeated this process 1,000 times and produced the bootstrapped 1,000 stacked profiles, with which the covariance between different radial bins was computed. In Fig.~\ref{fig:1stack-ksz}, the 1$\sigma$ statistical uncertainty is represented by the width of the lines, which is a square root of diagonal terms of the covariance matrix.

\subsection{Comparison with hydrodynamic simulations}
\label{subsec:sim}

\begin{figure*}
    \centering
    \includegraphics[width=0.33\linewidth]{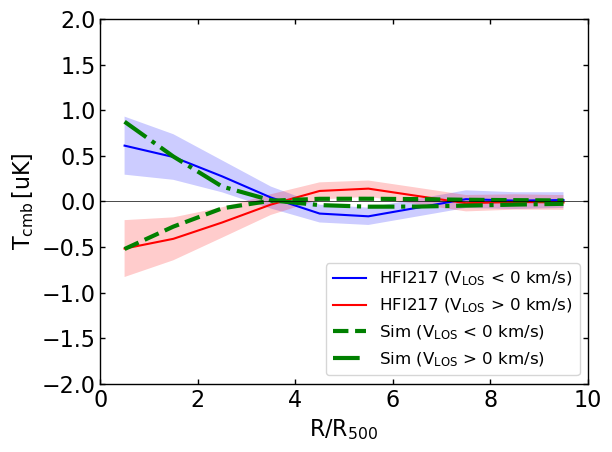}
    \includegraphics[width=0.33\linewidth]{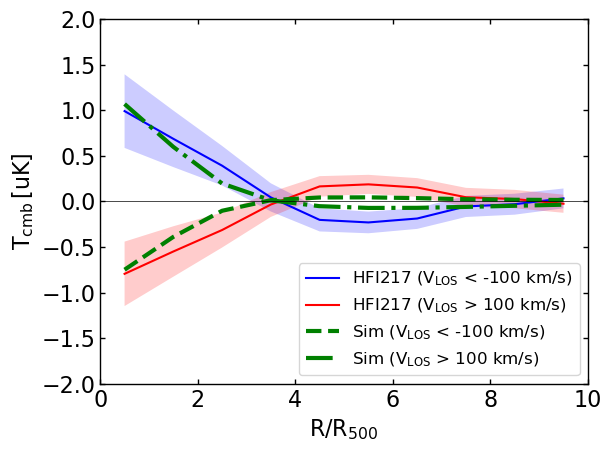}
    \includegraphics[width=0.33\linewidth]{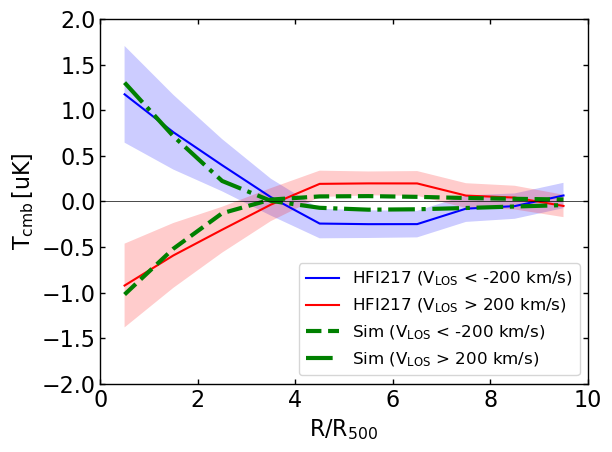}
    \caption{Positive kSZ radial profile ({\it blue}) around the 14,093 WHL clusters with the \planck\ temperature map at 217 GHz with the 1$\sigma$ uncertainty estimated by a bootstrap resampling, and negative kSZ radial profile ({\it red}) around the 16,338 WHL clusters, compared to positive kSZ radial profiles ({\it green dashed}) and negative kSZ radial profile ({\it green dash-dotted}) from the Magneticum hydrodynamic simulations. {\it Middle}: Same as {\it left} panel with an additional LOS velocity cut of 100 km/s. {\it Right}: Same as {\it left} panel with an additional LOS velocity cut of 200 km/s. }
    \label{fig:ksz-sim}
    \end{figure*}

We compared our measured kSZ signal with predictions from the Magneticum hydrodynamic simulation. 
For comparison, we performed the same stacking analysis with the simulated kSZ map and galaxy clusters from the Magneticum light-cone simulation as was performed with the real data. 
The mass and redshift distributions of the simulated galaxy clusters are matched to the real data. 
However, because a galaxy catalog is not provided in the Magneticum light-cone simulation, we added uncertainties to the LOS velocities of the simulated galaxy clusters randomly using a Gaussian function with the standard deviation of 260 km/s (see Sect. \ref{sec:vel}).
Based on the LOS velocities including the uncertainties, the simulated galaxy clusters were separated according to their positive or negative motions and stacked with the simulated kSZ map separately. 
(Some cancellation of the kSZ signal due to the uncertainties of the LOS velocities are included in the kSZ profile from the simulation as in the real data.)
The result is shown as a {\it green dashed line} in Fig.~\ref{fig:ksz-sim} and is compared to the data profile. In the left panel of Fig. 4, we can see that our measured kSZ profiles with positive and negative LOS motions are consistent with the ones from the simulation.
We also applied additional LOS velocity cuts of 100 km/s and 200 km/s to the simulated galaxy clusters in the middle and right panel, respectively. Our measured kSZ profiles show a similar amount of correlation between the amplitude of the kSZ signal and the LOS velocity to that predicted from the simulation.

\subsection{Null tests and significance}
\label{subsec:sig}

\begin{figure}
    \centering
    \includegraphics[width=\linewidth]{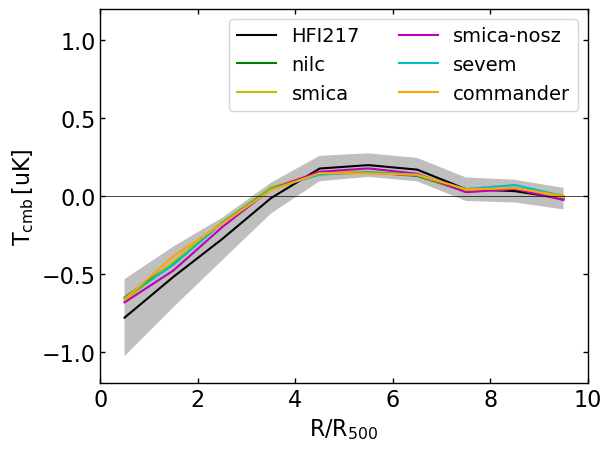}
    \caption{Velocity-weighted kSZ radial profile around the 30,431 WHL clusters with the \planck\ temperature map at 217 GHz ({\it black}) and with the 1$\sigma$ uncertainty estimated by a bootstrap resampling, compared to the kSZ radial profiles with the \planck\ CMB maps produced by different component separations of NILC ({\it green}), SMICA ({\it yellow}), SMICA-noSZ ({\it purple}), SEVEM ({\it cyan}), and COMMANDER ({\it orange}).}
    \label{fig:ksz-cmb}
    \end{figure}

To estimate the significance of the detected kSZ signal from the 30,431 WHL galaxy clusters, we stacked the filtered \planck\ maps at the positions of the clusters in the same way, as described above, with each cluster weighted by the LOS velocity as
\beq
T(R) = \frac{\sum_i T_i(R) \times  \varv_{i,\rm LOS} / \sigma_i^2}{\sum_i |\varv_{i,\rm LOS}| / \sigma_i^2},
\eeq
where $T_i(R)$ is the temperature value of the $i$-th cluster at the radial distance, $R$, $\varv_{i,\rm LOS}$ is the LOS velocity of the $i$-th cluster, and $\sigma_i$ is the variance of temperature values within the region we consider ($10 \times \theta_{500}$) centered on the $i$-th cluster. 
In this process, the kSZ signals with positive and negative LOS velocities end up with the same sign: a positive kSZ signal weighted by a negative LOS velocity has a negative signal and a negative kSZ signal weighted by a positive LOS velocity also has a negative signal. Therefore, the kSZ signals can be stacked without suffering any cancelation, while other components are canceled out. In addition, a cluster with a low LOS velocity (i.e., a weaker kSZ signal) is underweighted in the stacking. The stacked radial profile with the additional weight is shown in Fig.~\ref{fig:ksz-cmb} with the uncertainties estimated by bootstrap. 
An oscillating angular pattern is seen in the stacked radial profile. This is the result of the convolution of kSZ and CMB with the filter in Fig.~\ref{fig:cmb-filter}.  The pattern is also seen in the null-test profile in Fig.~\ref{1stack-null} and the $\beta$-model profile in Fig.~\ref{fig:fit}. We refer to this velocity-weighted kSZ profile from now on. 

First, we estimated the excess of the measured velocity-weighted kSZ profile to the null hypothesis. The signal-to-noise ratio (S/N) can be estimated with
\beq
S/N = \sqrt{\chi^2_{\rm data} - \chi^2_{\rm null}}
\label{eq:snr}
,\eeq
where
\beqa
\chi^2_{\rm data} &=& \sum_{i,j} T_{\rm data}(R_{i})^{T} (C_{ij}^{-1}) \, T_{\rm data}(R_{j}) ,\\
\chi^2_{\rm null} &=& \sum_{i,j} T_{\rm null}(R_{i})^{T} (C_{ij}^{-1}) \, T_{\rm null}(R_{j}),
\eeqa
where $T_{\rm data}(R_{i})$ is the temperature value at the $R_{i}$ bin of the data kSZ profile, $T_{\rm null}(R_{i})$ is the temperature value at the $R_{i}$ bin under the null hypothesis, which is zero, and $C_{ij}$ is the covariance matrix of the data profile, estimated by the bootstrap resampling. By measuring the kSZ signal up to $4\times\theta_{500}$, the S/N value was estimated to be 3.5 $\sigma$.

We also performed a Monte Carlo-based null test to assess the significance of our measurements. In the null test, we displaced the centers of the galaxy clusters at random positions on the sky and then the \planck\ maps were stacked at these random positions. We repeated this 1000 times to assess the $rms$ fluctuations of the foreground and background signals. The result shows that the average of the null-test profiles is consistent with zero, as shown in {\it cyan} in the left panel of Fig.~\ref{1stack-null} with the rms fluctuations, suggesting that our measurements are unbiased. The mean null-test profile shows the same coherent angular pattern as seen in the data profile in Fig.~\ref{fig:ksz-cmb}, which is due to the convolution of CMB with our filter in Fig.~\ref{fig:cmb-filter}.
Using the null-test sample, the significance of the measured kSZ signal can  also be estimated with Eq.\,\ref{eq:snr} by replacing $T_{\rm null}(R_{i})$ and $C_{ij}$ with their corresponding values from the null test. The S/N value was estimated to be 3.4$\sigma$ and is consistent with the result from bootstrap.

\section{Systematic errors}
\label{sec:systematics}

We performed one null test by displacing the galaxy clusters at random positions on the \planck\ maps in Sect. \ref{subsec:sig}. 
In this section, we describe two additional null tests that we performed to estimate potential systematic errors in our measurements. Subsequently, we checked the contamination in our measured kSZ signals due to the CMB, tSZ, and CIB, respectively.

\subsection{Additional null tests}
\begin{figure*}
    \centering
    \includegraphics[width=0.33\linewidth]{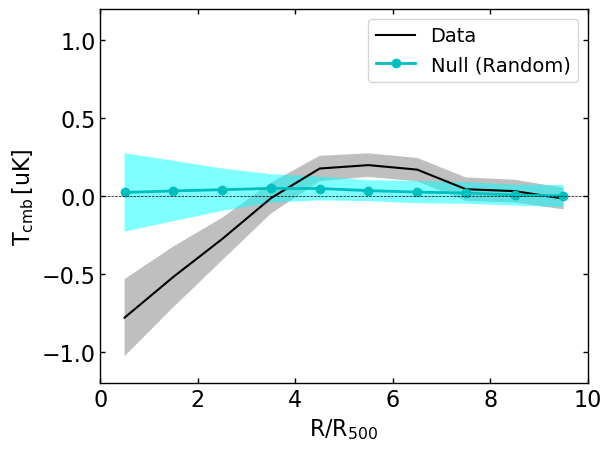}
    \includegraphics[width=0.33\linewidth]{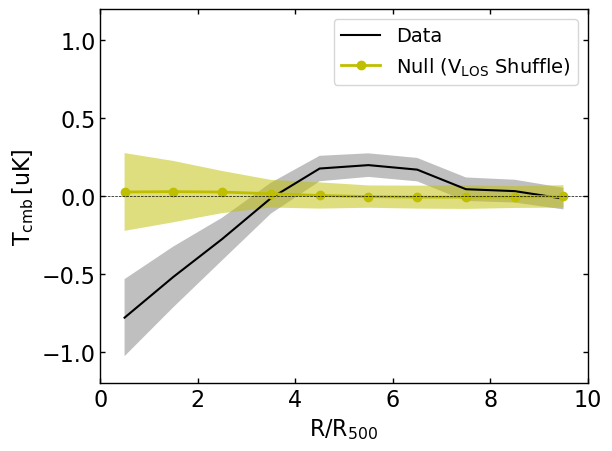}
    \includegraphics[width=0.33\linewidth]{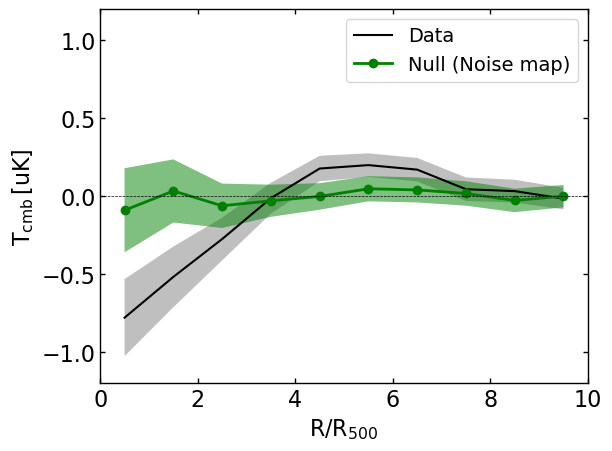}
    \caption{Velocity-weighted kSZ radial profile around the 30,431 WHL clusters with the \planck\ temperature map at 217 GHz ({\it black}) compared to three null tests. In the {\it left} panel, the clusters are displaced at random positions on the \planck\ map, and then stacked. This process is repeated 1,000 times and the mean of the 1,000 random samples is computed ({\it cyan}). The 1$\sigma$ uncertainty is estimated by computing a standard deviation of the 1,000 random samples. The  {\it middle} panel shows the results from cluster stacking after randomly shuffling the LOS velocities of the clusters; this process is repeated 1,000 times and the mean of the 100 velocity-shuffled profiles is computed ({\it yellow}). The 1$\sigma$ uncertainty is estimated by computing a standard deviation of the 1,000 velocity-shuffled profiles. In the {\it right} panel, the clusters are stacked with a noise map produced by $(T_{217}^{\rm HM1} - T_{217}^{\rm HM2})/2$, where $T_{217}^{\rm HM1(2)}$ is the half mission 1(2) \planck\ map at 217 GHz ({\it green}). The 1$\sigma$ uncertainty is estimated by a bootstrap resampling.}
    \label{1stack-null}
    \end{figure*}

For the first of the two additional null tests, we randomly shuffled the LOS velocities of the galaxy clusters, and then
the clusters were stacked with weights based on the shuffled LOS velocities.
This was done to test the effect of correlation between LOS velocities and
clusters. One shuffling may not be enough to remove the correlation, and therefore
we repeated the random velocity shuffling 1,000 times and evaluated the mean
and standard deviation of the 1,000 stacked profiles with shuffled velocities.
The second test consisted in performing the stacking with a noise map produced by $(T_{217}^{\rm HM1} - T_{217}^{\rm HM2})/2$, where $T_{217}^{\rm HM1(2)}$ is the half mission 1(2) \planck\ map at 217 GHz. The associated results are shown in the middle ({\it yellow}) and right ({\it green}) panels of Fig.~\ref{1stack-null} with the uncertainties. As expected, the obtained profiles are both consistent with zero. Including the null test performed in Sect. \ref{subsec:sig}, all three null tests suggest that our measurements of the kSZ effect are unbiased. 

\subsection{Contamination from Galactic and extragalactic emissions}
\label{subsec:contamination-all}

The kSZ signal is only a subdominant component and the measurement may be contaminated by Galactic as well as extragalactic emissions such as CMB, tSZ, and CIB. The CMB, the Galactic emissions, and instrumental noise are uncorrelated with cluster positions and are added as noise in our measurements. The tSZ and CIB are correlated with clusters, but it can be assumed that they are uncorrelated with their LOS velocities and canceled out in our analysis. No bias is expected that can be attributed to other components. 
For a further check, we compared the kSZ signal extracted with the \planck\ 217 GHz map to the kSZ signals extracted with the \planck\ CMB maps produced by different component-separation methods such as NILC, SMICA, SMICA-noSZ, SEVEM, and COMMANDER.  We note that the Galactic and extragalactic components are not cleaned in the \planck\ 217 GHz map, but are cleaned in the \planck\ CMB maps. The extracted kSZ signals are all consistent, as shown in Fig.~\ref{fig:ksz-cmb}, regardless of the maps we analyze, with or without the Galactic and extragalactic components. This result suggests that the contamination of our measurements from the Galactic and extragalactic components is minor. 

\subsection{Contamination of the CMB}
\label{subsec:contamination-cmb}

While we have shown that the contamination of our measured kSZ signals from the Galactic and extragalactic components is minor, contamination from the CMB may still be present at some level. This is because the null tests that we performed involving random displacements and velocity shuffling of galaxy clusters show no statistical bias as 1,000 realizations are considered, but the CMB is one realization and the cancelation by stacking may not be sufficient for the real CMB.  Therefore, we estimated a possible level of contamination only from the CMB by simulating 100 CMB maps with different realizations using the PYCAMB\footnote{https://github.com/steven-murray/pycamb} interface to the CAMB\footnote{http://camb.info} code \citep{Lewis2000}. We repeated our stacking analysis for the 100  simulated CMB maps, producing 100 CMB profiles. The mean of the 100 CMB profiles is consistent with zero and its standard deviation is $\sim23\%$ of the amplitude of our measured kSZ signal. Thus, while the CMB contamination may be present at some level, it should not be significant. 

\subsection{Contamination of the tSZ}
\label{subsec:contamination-tsz}

The contamination of the tSZ can also be estimated quantitatively by applying our stacking analysis to the \planck\ all-sky $y$ maps provided in the \planck\ 2015 data release\footnote{https://pla.esac.esa.int} \citep{Planck2016XXII}. We used the $y$ map from the modified internal linear combination algorithm (MILCA) \citep{hurier2013}, but the result is consistent using the $y$ map from needlet independent linear combination (NILC) \citep{remazeilles2013}. The Compton $y$ parameter in the \planck\ HFI 217 GHz map can be calculated with its frequency dependence, given by
\beq
\frac{\Delta{T}_{\rm tsz}}{T_{\rm CMB}} = g(\nu) \, y, 
\label{eq:tsz}
\eeq
where $g(\nu) = x \coth(x/2)-4$ with $x = h \nu/(k_{\rm B} \, T_{\rm CMB})$, where $h$ is the Planck constant, $k_{\rm B}$ is the Boltzmann constant, and $T_{\rm CMB}$ is the temperature of the CMB. At 217 GHz, $T_{\rm CMB} \,  g(\nu) = 0.187$ K$_{\rm CMB}$ is given in \cite{Planck2016XXII} for the conversion from the Compton $y$ parameter to CMB temperature. The result of stacking the $y$ map shows that the tSZ contamination is $\sim$1\% of the amplitude of our measured kSZ signal and is thus negligible.  

\subsection{Contamination of the CIB}
\label{subsec:contamination-cib}

Similarly, we estimated the level of contamination from the CIB by stacking the \planck\ all-sky CIB maps provided in \cite{Planck2016IRXLVIII}. However, as the CIB map at 217 GHz was not produced, we scaled the \planck\ CIB map at 353 GHz to the one at 217 GHz with the power-law spectral index of $\beta\sim1.1$ \citep{Tucci2016}. The result of stacking the CIB map shows that the CIB contamination is $\sim6\%$ of the amplitude of our measured kSZ signal and not significant. 

\section{Interpretation}
\label{sec:interp}

The CMB temperature fluctuation caused by the kSZ effect is given by
\beq
\frac{\Delta{T}_{\rm ksz}}{T_{\rm CMB}} = - \sigma_{\rm T} \int n_{\rm e} \, \left( \frac{\bm{\varv}\cdot \bm{\hat{n}}}{c}  \right) \, \der{l} \simeq - \tau \left( \frac{\bm{\varv} \cdot \bm{\hat{n}}}{c} \right)
\label{eq:ksz}
,\eeq
where $\sigma_{\rm T}$ is the Thomson scattering cross section, $c$ is the speed of light, $n_{\rm e}$ is the electron number density, and $\bm{v} \cdot \bm{\hat{n}}$ represents the peculiar velocity of electrons along the line of sight. In the final transformation, the integral, $\tau = \sigma_{\rm T} \int n_{\rm e} \der{l} $, was performed along the line of sight under the approximation that the typical correlation length of LOS velocities (given by $\bm{\varv} \cdot \bm{\hat{n}}$) is much larger than the density correlation length, and therefore the LOS velocity term can be extracted from the kSZ integral. This is justified by \cite{Planck2016IRXXXVII} who find that the typical correlation length of peculiar velocities is 80–100 \mpc, well above the typical galaxy correlation length of $\sim$5\mpc. 

Physical properties of gas can be estimated by considering a $\beta$ model \citep{Cavaliere1978} for a gas(electron) density profile, given by 
\beq
n_{\rm e}(r) = n_{\rm e,0} \left( 1+\frac{r}{r_{\rm c}}\right)^{-3\beta/2}, 
\label{eq:n3d}
\eeq
where $n_{\rm e,0}$ is the central electron density, $r$ is the cluster radial extension, and $r_{\rm c}$ is the core radius of the electron distribution. In our study, we used $\beta=0.86$ and $r_{\rm c} = 0.2 \times R_{500}$ from the measurements of the South Pole Telescope clusters \citep{Plagge2010}. We can express the data profile as a geometrical projection of the density profile with $n_{\rm e}(r)$, 
\beq
\tau(R) = \sigma_{\rm T} \int \frac{2r \, n_{\rm e}(r)}{\sqrt{r^2 - R^2}} \, \der r, 
\label{eq:n2d}
\eeq
where $R$ is the tangential distance from a galaxy cluster. (We represent the 3D distance with the lowercase letter $r$, and the 2D distance on a map with the uppercase letter $R$.) 

\begin{figure}
    \centering
    \includegraphics[width=\linewidth]{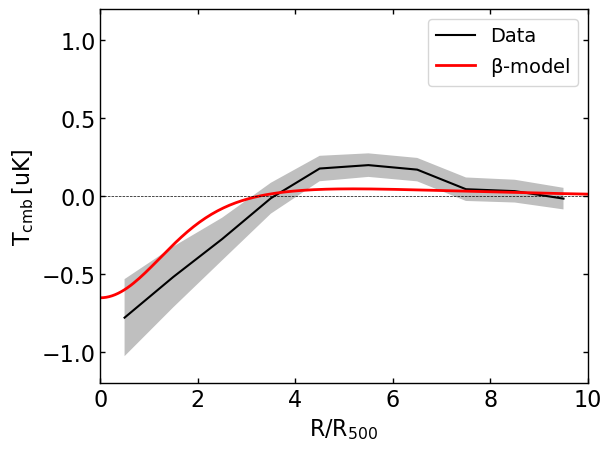}
    \caption{Velocity-weighted kSZ radial profile around the 30,431 WHL clusters with the \planck\ temperature map at 217 GHz ({\it black}), fitted with the $\beta$ model ({\it red}). }
    \label{fig:fit}
    \end{figure}

We fit this model to the measured kSZ profile using the average of the LOS velocities (in absolute value) of the WHL galaxy clusters estimated in Sect. \ref{sec:vel}. 
However, the average velocity is overestimated because of the uncertainties of the LOS velocities. (The Gaussian distribution of the estimated LOS velocities have a larger standard deviation than the distribution of the true LOS velocities because of the uncertainties.) The average velocity can be corrected with the uncertainty of the LOS velocities investigated in Sect. \ref{sec:vel}. The uncertainty on the LOS velocities also induces the decrease in the amplitude of the measured kSZ signal \citep{Nguyen2020} because it causes some cancelation of the kSZ signal. This can be corrected analytically with the uncertainty on the LOS velocities estimated in Sect. \ref{sec:vel}. Including these corrections in the model, we fit this 
$\beta$  model to the data. 
The result of the model fitting is shown in Fig.~\ref{fig:fit}. The reduced $\chi^2$ value is 1.3. We note that we can see a coherent angular pattern in the model profile similar to the data profile, which is due to the convolution of the $\beta$ profile with our filter in Fig.~\ref{fig:cmb-filter}.
Defining the optical depth of a galaxy cluster within $R_{500}$ as 
\beq
\tau_{\rm e, 500} = \int_{0}^{R_{500}} \sigma_{\rm T} \, n_{\rm e}(r) \, \der{V}, 
\eeq
the fitting result provides the average optical depth of the WHL clusters:
\beq
\overline{\tau}_{\rm e, 500} =  (3.0 \pm 0.9) \times 10^{-3}. 
\label{eq:tau-measure}
\eeq

Offsets between cluster centers from optical data and centers of gas distribution in clusters may have an impact on the estimated optical depth. \cite{Rozo2014} studied the offsets between X-ray cluster centers (proxy of gas center) and optical cluster centers and showed that $\sim70\%$ of the WHL clusters have offsets of less than $\sim 0.1$ Mpc from X-ray clusters. This latter distance corresponds to $\sim 0.2$ arcmin at the median redshift of our sample ($z \sim 0.44$), which is smaller than the angular resolution of the \planck\ maps ($\sim 5$ arcmin). To check the contribution of this offset to the measurement of the optical depth, we randomly displaced the centers of the WHL clusters by distances drawn from a Gaussian distribution with the standard deviation of $\sim 0.1$ Mpc and repeated the stacking. The result shows that the amplitude of our measured kSZ profile decreases only by $\sim 3 \%$ and the effect is minor.

We can also estimate a total gas mass in a galaxy cluster defined as 
\beq
M_{gas,500} = \int_{0}^{R_{500}} n_{\rm e}(r) \, \mu_{\rm e} \, m_{\rm p} \, \der{V}, 
\label{eq:mgas}
\eeq
where $\mu_{\rm e} =  1.148$ is the mean molecular weight of electrons \citep{Arnaud2010}, and $m_{\rm p}$ is the mass of proton. Through our measurement of the optical depth, the average gas mass in the WHL clusters can be estimated to be $\overline{M}_{gas,500} \sim 1.2 \times 10^{13}$ \msun. This provides a gas mass fraction of $f_{gas,500} = M_{gas,500}/M_{500} = 0.12 \pm 0.04$ for our sample with an average mass of $M_{500} \sim 1.0 \times 10^{14}$ \msun.

The Magneticum simulation also provides gas masses of the simulated clusters within $R_{500}$, and reveals a gas mass of $M_{gas,500} \sim 1.3 \times 10^{13}$ \msun\ for the cluster with $M_{500} \sim 1.0 \times 10^{14}$ \msun\ (average mass of our cluster sample). This corresponds to a gas mass fraction of $f_{gas,500} \sim 0.13$ and is consistent with our result.

\section{Discussion}
\label{sec:discussion}

The kSZ effect is sensitive to cosmological parameters such as the growth rate of density perturbations. 
However, we find that our result is consistent using the \wmap\ and \planck\ cosmology and the current level of our kSZ detection does not allow us to constrain cosmological parameters. 
We therefore used our measurement of the kSZ signal to constrain the average optical depth (or gas mass) of the cluster sample. Our result is compared to other measurements as follows. 

Gas masses in groups and clusters of galaxies were studied in X-rays by \cite{Gonzalez2013}.  The authors combined measurements with \xmm\ and \chandra\ observations from \cite{Vikhlinin2006}, \cite{Sun2009}, and \cite{Sanderson2013}, and showed the X-ray gas fraction within $R_{500}$ as a function of $M_{500}$. The relation shows $f_{gas,500} \sim 0.1$ at $M_{500} \sim 1.3 \times 10^{14}$ \ms\ (average mass of our sample), which is slightly lower than our result of $f_{gas,500} = 0.12 \pm 0.04$, but is consistent within our measurement uncertainties. 

Gas in halos was studied with the kSZ in \cite{Planck2016IRXXXVII}. The authors detected the pairwise kSZ signal at the positions of central galaxies from the SDSS DR7 data using the \planck\ and \wmap\ maps and measured an average optical depth of $\overline{\tau} =  (1.4 \pm 0.5) \times 10^{-4}$. This value is one order of magnitude lower then our value of $\overline{\tau}_{\rm e, 500} =  (3.0 \pm 0.9) \times 10^{-3}$, but this discrepancy can be attributed to the difference in halo mass: the halo mass of their sample is $M_{500} \sim 0.3 \times 10^{14}$ \ms, which is significantly lower than the average mass of our sample, $M_{500} \sim 1.3 \times 10^{14}$ \ms.

A similar study to that of \cite{Planck2016IRXXXVII} was performed with the kSZ by \cite{Soergel2016} using clusters with photometric redshifts. The authors detected the pairwise kSZ signal by combining a cluster catalog from the Dark Energy Survey (DES) with the CMB map from the South Pole Telescope Sunyaev-Zel'dovich Survery (SPT-SZ) and measured an average optical depth of $\overline{\tau} = (3.75 \pm 0.89) \times 10^{-3}$ for the clusters with $M_{500} \sim (1-3) \times 10^{14}$ \ms. This value is higher than our value of $\overline{\tau}_{\rm e, 500} = (3.0 \pm 0.9) \times 10^{-3}$, but this discrepancy may again  be related to the difference in mass: the average mass of our cluster sample, $M_{500} \sim 1.3 \times 10^{14}$ \ms, corresponds to the lower end of their cluster mass range.  \cite{Soergel2016}  also estimated the gas mass fraction within $R_{500}$ to be $f_{gas,500} = 0.08 \pm 0.02$ assuming a $\beta$ model for the density distribution. Our value of $f_{gas,500} = 0.12 \pm 0.04$ is slightly higher than their result, but consistent within $\sim 1 \sigma$ when we consider the uncertainties on both measurements. 

\cite{Lim2020} also detected the kSZ signals from galaxy groups and clusters in the mass range of $2 \times 10^{12}$ \ms $< M_{500} <$ $2 \times 10^{14}$ \ms\ by combining the cluster catalog from \cite{Yang2007} and the \planck\ frequency maps at 100, 143, and 217 GHz. Surprisingly, their results show that the gas fraction in halos is consistent with the universal baryon fraction in their cluster mass range ($f_{gas} \sim 0.17$). To compare our results with theirs, we converted our measurements to their estimator of $\widetilde{K}_{500}$, which is the intrinsic kSZ signal scaled to redshift $z$ = 0 and to a fixed angular diameter distance. Our value translates to $\widetilde{K}_{500} = (1.4 \pm 0.4) \times 10^{-2}$ at $M_{500} \sim 1.3 \times 10^{14}$ \ms, and the value is consistent with their value within $\sim1 \sigma$.

\cite{Lim2020} claim that the gas fraction in halos is consistent with the universal baryon fraction down to the mass of $2 \times 10^{12}$ \ms, but this result is not consistent with those of \cite{Soergel2016} or the X-ray measurements by \cite{Gonzalez2013}, showing a lower gas fraction than the universal value. One possible explanation for this discrepancy may be because of the filter used in \cite{Lim2020} to extract the signals. Indeed, \cite{Lim2020} used a matched filter to extract the kSZ signal, which may induce a bias, if the assumed profile is incorrect \citep{Ferraro2015}. However, given the large \planck\ beam, they did not find a significant bias in their results by testing with an incorrect profile. In addition, while they assumed the $\beta$ profile from \cite{Plagge2010}, we used the same $\beta$ profile and find that it closely fits our measured kSZ profiles with the reduced $\chi^2$ of 1.3. Therefore, the assumed profile does not explain the difference. On the other hand, our measured kSZ profiles show extended kSZ signal beyond $R_{500}$ and this signal does not seem to be consistent with the result of \cite{Lim2020}  that the gas fraction within $R_{500}$ of halos is approximately equal to the universal baryon fraction. However, the beam of the \planck\ maps ($\sim 5$ arcmin) is equivalent to the average angular size of our cluster sample ($\sim 4$ arcmin) and this coarse beam prevents a definitive conclusion. Another possible explanation for this discrepancy may be due to the difference in cluster redshifts. \cite{Lim2020} studied a cluster sample at $z$ < 0.12, while the sample in \cite{Soergel2016} is at $z \sim 0.5$. Thus, the evolution of the gas in halos may explain the difference in the kSZ measurements. However, the X-ray measurement in \cite{Gonzalez2013} was also applied to local clusters at $z\sim0.1$ and the evolution does not seem to explain the difference. So far, the reason for the discrepancy is unknown. Our result is statistically consistent with \cite{Lim2020} as well as \cite{Soergel2016} and \cite{Gonzalez2013}, and more precise measurements of gas are needed in order to come to firm conclusions. 

\section{Conclusion}
\label{sec:conclusion}

In this paper, we present the first direct detection of the kSZ signal with a significance of 3.5$\sigma$. 
The measurement was performed by stacking the \planck\ temperature map at 217 GHz from the \planck\ 2018 data release at the positions of Wen-Han-Liu (WHL) galaxy clusters constructed from the SDSS galaxies. 
If all the clusters are simply stacked, the kSZ signals are canceled out because of the equal probability of clusters showing positive or negative kSZ signals. 
To avoid this cancelation, we estimated the peculiar velocities of the galaxy clusters along the LOS through the galaxy density field computed from the SDSS galaxies, which is related to the matter density field with the galaxy bias of $b = 2$ estimated in other studies. 
Using the LOS velocities as a weight in the stacking, the positive and negative kSZ signals are turned to the same sign and are added together, while other components are canceled out by the positive or negative LOS velocity with equal probability.
The measured kSZ signals show a clear correlation with the amplitude of the LOS velocities. As a result of the stacking, we obtain an average kSZ profile for the galaxy clusters with a mass of $M_{500} \sim 1.0 \times 10^{14}$ \msun, showing an extended distribution of gas around the galaxy clusters beyond $3 \times R_{500}$. 

The kSZ signal is a subdominant component and our measurement may be contaminated by Galactic and extragalactic emission. The CMB, the Galactic emission, and instrumental noise are uncorrelated with cluster positions and are added as noise in our measurements. The tSZ and CIB are correlated with clusters, but uncorrelated with their LOS velocities.  We performed three different null tests, the results of which are consistent with zero, indicating that our measurements are unbiased. Possible contamination from the CMB, tSZ, and CIB was also investigated but we did not find any significant contamination from these components. 

Based on our kSZ measurement, we estimated the average optical depth and found $\overline{\tau}_{\rm e, 500} = (3.0 \pm 0.9) \times 10^{-3}$ for clusters with the mass of $M_{500} \sim 1.0 \times 10^{14}$ \msun\ assuming a $\beta$ model. This provides an average gas mass of the galaxy clusters of $\overline{M}_{gas,500} \sim 1.2 \times 10^{13}$ \msun, leading to a gas fraction of $f_{gas,500} = 0.12 \pm 0.04$ within $R_{500}$. 
We compared our measurement of gas fraction with that calculated from the Magneticum hydrodynamic simulations as well as the measurements from X-rays \citep{Gonzalez2013} and kSZ \citep{Soergel2016}, which show that the gas mass fraction is lower than the cosmic baryon fraction. We also compared our result with that of \cite{Lim2020}, who, on the other hand, claim that the gas fraction in halos is approximately equal to the universal baryon fraction down to the mass of $2 \times 10^{12}$ \ms, which is in disagreement with the measurements by \cite{Gonzalez2013} and \cite{Soergel2016}. Our result is statistically consistent with the results of these authors, and more precise measurements of gas are necessary in order to make firm conclusions. 


\appendix

\section{Components at HFI 217 GHz}
\label{sec:comp}

The stacking of the \planck\ temperature map at 217 GHz shows an excess at the positions of the WHL clusters, as seen for example in the black line in the upper left panel of Fig.~\ref{fig:1stack-ksz}. We investigated the contribution to this excess from the Galactic and extragalactic components. 

In the Galactic emissions, we considered the dust and CO emission. The other Galactic emissions such as synchrotron are minor at 217 GHz according to \cite{Planck2016X}. The all-sky Galactic dust map was provided in \cite{Planck2016IRXLVIII}. However, as the dust map at 217 GHz was not produced, we scaled the \planck\ dust map at 353 GHz to the one at 217 GHz with the power-law spectral index of $\beta\sim1.5$ \citep{Planck2016X}. The CO emission map at 230 GHz, because of the CO rotational transition of J=2-1, was produced in \cite{Planck2016X}. In the extragalactic emissions, we consider the tSZ and CIB component and use the tSZ and CIB maps scaled at 217 GHz, as described in Sect. \ref{subsec:contamination-tsz} and Sect. \ref{subsec:contamination-cib}. 

\begin{figure}[ht!]
    \centering
    \includegraphics[width=\linewidth]{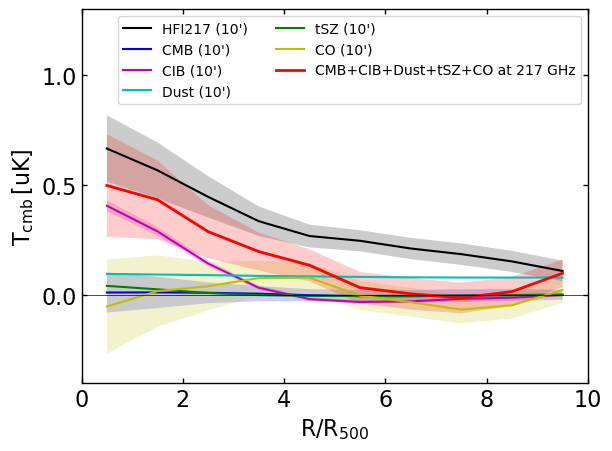}
    \caption{Radial profile around the 30,431 WHL clusters with the \planck\ temperature map at 217 GHz ({\it black}) with contributions from CMB ({\it blue}), CIB ({\it magenta}), the Galactic dust ({\it cyan}), tSZ ({\it green}), and CO emissions ({\it yellow}) as well as the total ({\it red}).  }
    \label{fig:comps}
    \end{figure}

We stacked these maps at the positions of the WHL clusters and compare the stacked signals in Fig.~\ref{fig:comps}. We note that all these maps were convolved by an additional beam, giving rise to a common beam size of 10 arcmnin in FWHM, which is the largest beam of the \planck\ tSZ map. The dominant contribution at small scales is from the CIB and that from the rest of the components is minor. The signals at large scales cannot be well explained  by the components we consider. This may be due to an unknown spectral index of CIB or relatively uncertain CO emission. However, the signals at large scales do not contribute to our kSZ signals at small scales (<$4 \times R_{500}$) and we do not take the difference into account in this paper. 

\section{Contamination to the kSZ signal}
\label{sec:contam}

\begin{figure}
    \centering
    \includegraphics[width=\linewidth]{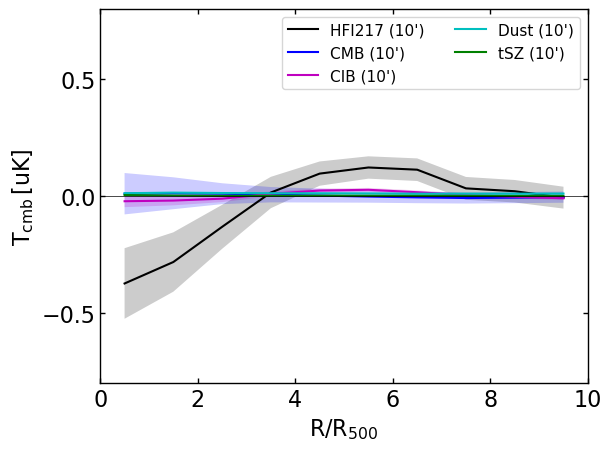}
    \caption{Contamination in our measured kSZ profile ({\it black}) due to contributions from CMB ({\it blue}), CIB ({\it magenta}), the Galactic dust ({\it cyan}), and tSZ ({\it green}).  }
    \label{fig:contam}
    \end{figure}

In order to evaluate the contamination of the measured kSZ signal attributable to the Galactic and extragalactic emissions, we used the maps of CMB, CIB, tSZ, and Galactic dust convolved to a common beam size of 10 arcmnin in FWHM, as described in Appendix. \ref{sec:comp}. We repeated the same stacking analysis in Sect. \ref{sec:ana} for these maps as for the HFI map at 217 GHz. The results are shown in Fig.~\ref{fig:contam} and the contribution from each component is summarized in Sect. \ref{sec:systematics}.

\begin{acknowledgements}
The authors thank an anonymous referee for the useful comments and suggestions. This research has been supported by the funding for the ByoPiC project from the European Research Council (ERC) under the European Union's Horizon 2020 research and innovation programme grant agreement ERC-2015-AdG 695561. The authors acknowledge fruitful discussions with the members of the ByoPiC project (https://byopic.eu/team). Furthermore, S.Z. acknowledge support by the Israel Science Foundation (grant no. 255/18).
We thank LSS2LSS science program of the $\Psi$2 initiative of the Institut Pascal, funded by the Paris-Saclay University, in which this research has been initiated.
This publication used observations obtained with \planck (\url{http://www.esa.int/Planck}), an ESA science mission with instruments and contributions directly funded by ESA Member States, NASA, and Canada. The authors thank Klaus Dolag and Antonio Ragagnin for providing the Magneticum simulations, and Adi Nusser for fruitfull discussions.
\end{acknowledgements}
\bibliographystyle{aa} 
\bibliography{ksz} 

\end{document}